\documentclass[aps,twocolumn,showpacs,preprintnumbers,amsmath,amssymb,prl]{revtex4-1}

\usepackage[colorlinks=true,bookmarks=false,citecolor=blue,urlcolor=blue]{hyperref} 

\usepackage{amsmath,amssymb}
\usepackage{graphicx}
\usepackage{verbatim} 
\usepackage{color}

\usepackage[english]{babel}
\usepackage{bm}
\usepackage{natbib}
\usepackage{graphicx}

\def\dfrac{\displaystyle\frac}  

\newcommand{\nix}[1]{}

\begin{document}
\title{Tunable nonlinear graphene metasurfaces}

\author{Daria A. Smirnova$^{1}$}
\author{Andrey E. Miroshnichenko$^{1}$}
\author{Yuri S. Kivshar$^{1}$}
\author{Alexander B. Khanikaev$^{2,3}$}\email[]{akhanikaev@qc.cuny.edu}

\affiliation{
$^{1}$~Nonlinear Physics Centre, Australian National University, Canberra ACT 2601, Australia \\
$^{2}$~Department of Physics, Queens College of The City University of New York, Queens, NY 11367, USA \\
$^{3}$~Department of Physics, The Graduate Center of The City University of New York, NY 10016, USA
}

\begin{abstract}
We introduce the concept of nonlinear graphene metasurfaces employing the controllable interaction between a graphene layer and a planar metamaterial. Such hybrid metasurfaces support two types of subradiant resonant modes, asymmetric modes of structured metamaterial elements ("metamolecules") and graphene plasmons exhibiting strong mutual coupling and avoided dispersion crossing. High tunability of graphene plasmons facilitates strong interaction between the subradiant modes, modifying the spectral position and lifetime of the associated Fano resonances. We demonstrate that strong resonant interaction, combined with the subwavelength localization of plasmons, leads to the enhanced nonlinear response and high efficiency of the second-harmonic generation.
\end{abstract}

\maketitle

{\em Introduction.} Unique electromagnetic properties of graphene are paving a way towards their applications in novel optical devices~\cite{1,2,3,4,5,6,7,8,9}. One of the main benefits of using graphene stems from the ability to control its optical response by changing the chemical potential via electrostatic or chemical doping~\cite{10,11,12,13,14}. While the electromagnetic response of bare graphene is considered to be strong, it is still not sufficient for a majority of practical applications. One of the promising approaches to enhance optical response of graphene is to place it into close proximity with metallic and dielectric photonic structures, and this approach  has been employed to design highly tunable infrared and THz metasurfaces enabling both amplitude and phase modulation~\cite{15,16,17,18,20,21,211}. In addition, doped graphene exhibits a plasmonic response, and therefore it can be patterned to form tunable planar plasmonic metasurfaces operating at infrared and THz frequencies~\cite{9,22,23}.

In this Letter, we discuss a novel type of Fano-resonant hybrid metasurfaces composed of a graphene layer and a planar metamaterial where two types of subradiant modes of different origin, asymmetric plasmonic modes of metallic metamolecules and graphene plasmons, are strongly coupled. We show that such metasurfaces support {\em cascaded Fano resonances}~\cite{27,28} originating from the interference of subradiant plasmonic modes in graphene and localized modes of metamolecules. We demonstrate that such hybrid graphene metasurfaces provide strong tunability and dramatic field enhancement giving rise to enhanced nonlinear response.

\begin{figure}[t!]
\centering\includegraphics[width=1.0\linewidth]{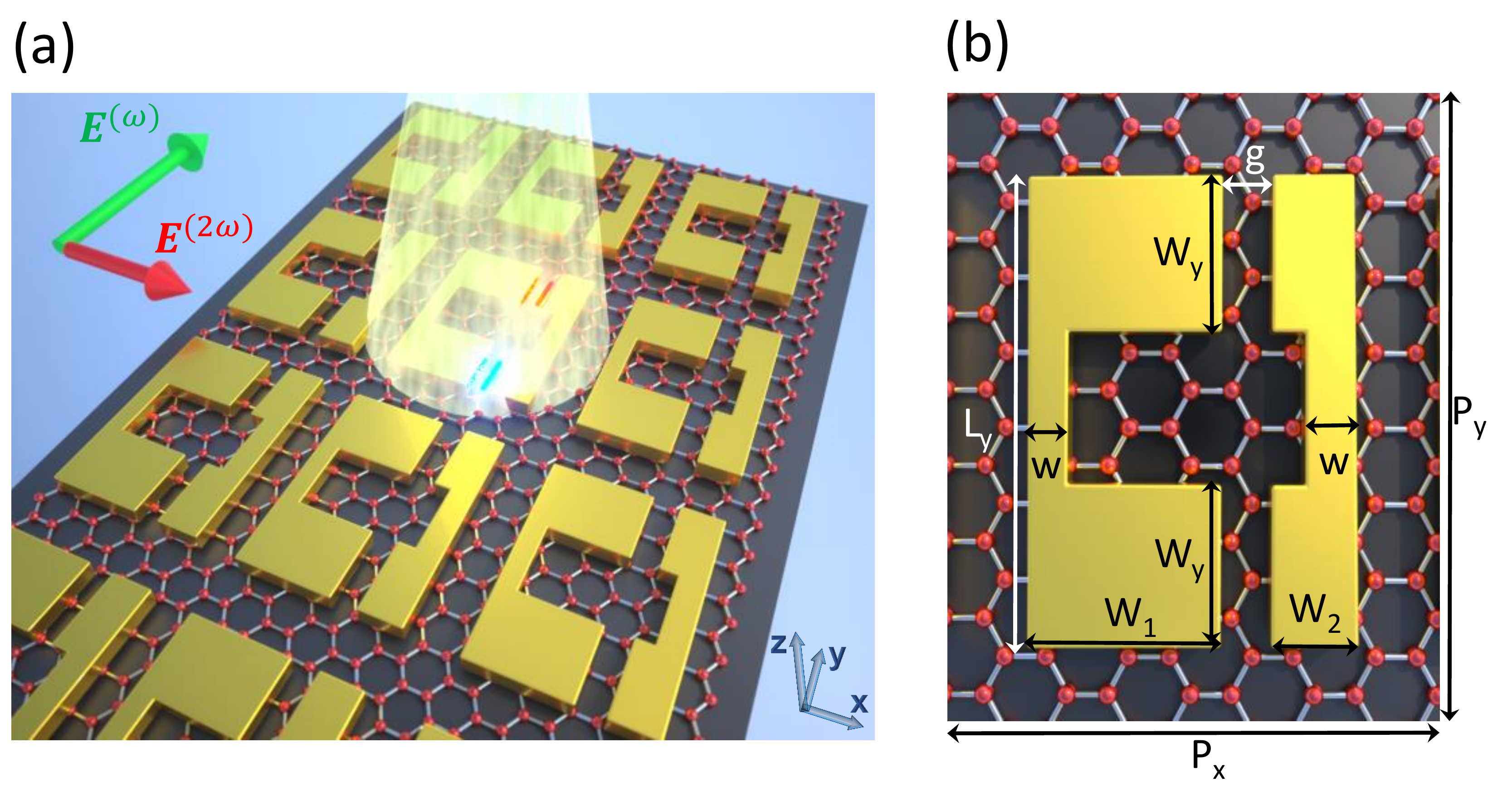}
\caption{(Color online) (a) Artistic view of the light interaction with a hybrid metasurface created by a lattice of asymmetric gold SRRs placed on top of a graphene layer. (b) Top view and size definitions of a unit cell of the hybrid metasurface.
}
\label{fig:fig1_1}
\end{figure}

{\em Model and parameters.} We consider a metasurface composed of a square lattice of metallic split-ring-resonators (SRRs) placed on top of a graphene layer, as shown in Figs.~\ref{fig:fig1_1}(a,b). The electromagnetic response is calculated by using the finite-element method solver COMSOL Multiphysics with a graphene layer modelled as a surface current. At low frequencies $\hbar \omega <2 \mathcal{E}_{F}$,  graphene is described by a frequency-dependent surface conductivity~\cite{26}
\begin{equation}
\sigma(\omega) = \displaystyle{\frac{ie^2}{\pi \hbar}
\left\{\frac{\mathcal{E}_{F}}{\hbar\left(\omega + i\tau_{\text{intra}}^{-1}\right)} + \frac{1}{4} \text{ln}\left|\frac{2\mathcal{E}_{F} -\hbar \omega}{2\mathcal{E}_{F} + \hbar \omega}\right|\right\}}\:,
\label{eq:sigma}
\end{equation}
where $e=-|e|$ is electron's charge, $\mathcal{E}_{F} = \hbar V_{F} \sqrt{\pi n}$ is the Fermi energy, $n$ is the doping electron density, $V_{F}\approx c/300$ is the Fermi velocity, and  $\tau_{\text{intra}} = 0.15$  ps is the relaxation time. At frequencies $\hbar \omega < \mathcal{E}_{F}$, the Drude-like conductivity originating in intra-band transitions dominates and graphene supports highly confined extremely short-wavelength p-polarized plasmons, whose dispersion can be tuned by changing the Fermi level of graphene $\mathcal{E}_F$. Considering the normal incidence of a $y$-polarized plane wave, we calculate numerically both transmission and reflection spectra for different values of the chemical potential.

\begin{figure}[t!]
\centering\includegraphics[width=0.9\linewidth]{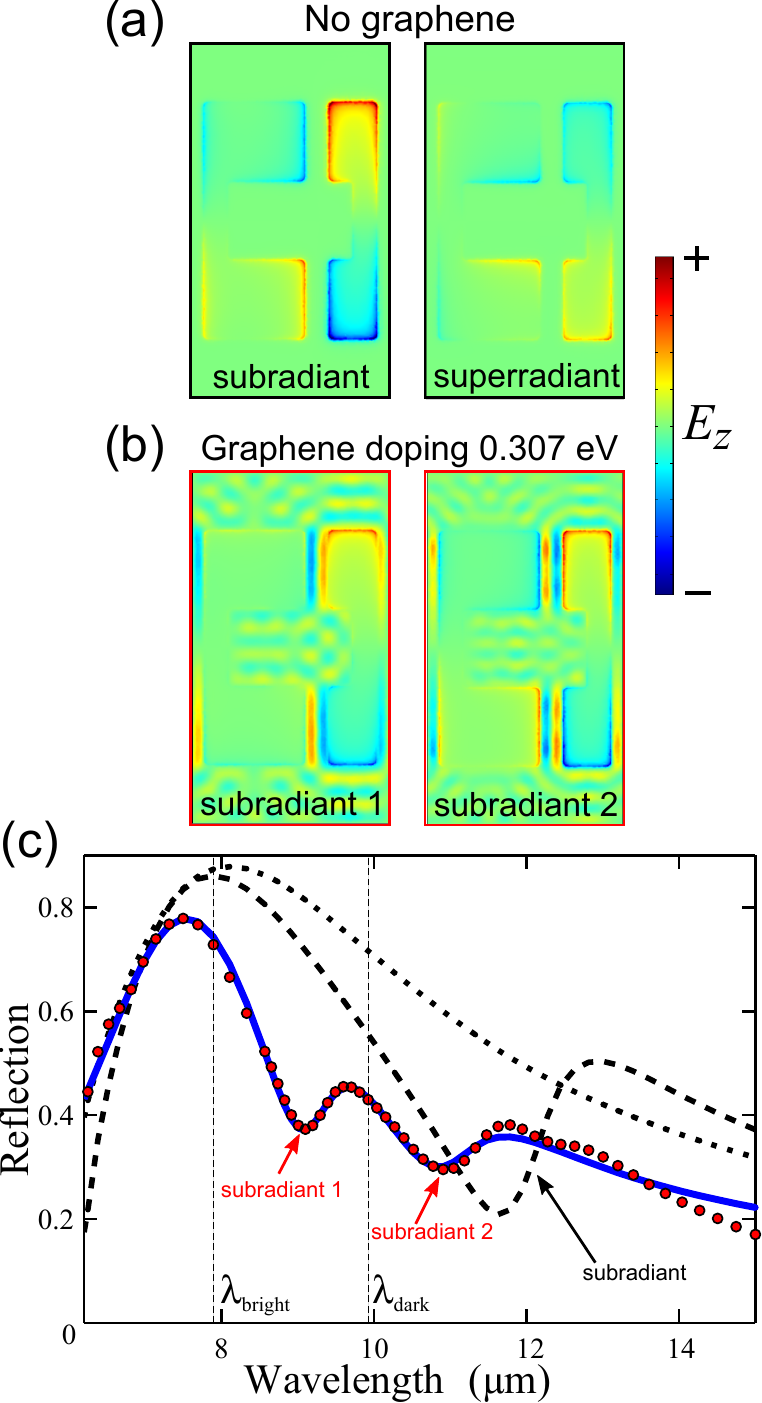}
\caption{(Color online) (a,b) Near-field images of the resonances shown by the $E_z$ field component without and with a graphene layer, respectively. (c) Numerically calculated reflection spectrum at $\mathcal{E}_F=0.307$ eV (red circles) along with the fitting obtained by the coupled-mode theory (blue solid curve). For the cases of symmetric and asymmetric split-rings, the spectra without graphene are shown by black dotted and dashed lines, respectively. Structure parameters are: the substrate permittivity $\varepsilon_2=1.35^2$, $P_x=2.5$ $\mu$m and $P_y=4.5$ $\mu$m, the parameters of the gold SRR are $L_y=3$ $\mu$m, $g=310$ nm, w$=0.3$ $\mu$m, $W_y=1$ $\mu$m, $W_1=1.28$ $\mu$m, $W_2=0.6$ $\mu$m, and gold thickness is $h=40$ nm.
}
\label{fig:fig1}
\end{figure}

Without a graphene layer, the metasurface becomes a conventional two-dimensional metamaterial of double SRRs, supporting symmetric (superradiant) and asymmetric (subradiant) modes that correspond to the in-phase and out-of-phase current distribution along two arms of the metamolecule [see Fig.~\ref{fig:fig1}(a)]. When the symmetry of the metamolecules is reduced, the interference between these two modes results in a characteristic Fano-like asymmetric lineshape in far-field spectra, the transition illustrated by black dotted and dashed lines in Fig.~\ref{fig:fig1}(c)~\cite{29,30,31,32}. By introducing a graphene layer, we can excite graphene plasmons whose wavelength can be tuned to match one of the characteristic scales of the metasurface. Note, however, that even without plasmonic modes of graphene tuned to the spectral range of interest, the surface conductivity of graphene modifies electromagnetic interactions on the metasurface leading to the overall blue shift of the resonances~\cite{16}.

\begin{figure}[t]
\centering\includegraphics[width=1\linewidth]{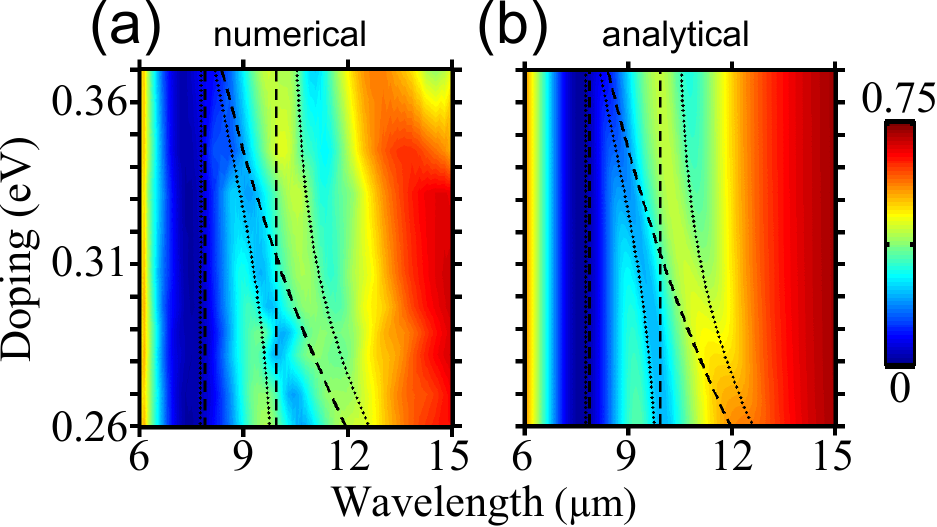}
\caption{(Color online) (a) Numerically calculated transmission spectra of the metasurface compared to (b) analytical results.
Black dotted curves in panels (a,b) 
illustrate the anticrossing between the dark modes.}
\label{fig:fig2}
\end{figure}

Due to the subwavelength nature of graphene plasmons, it is natural to match their wavelength to the smallest scale associated with the gaps of the split-ring resonators, which effectively become a cavity for plasmons~\cite{18}. Here we are interested in graphene plasmons forming a subradiant mode, and such mode can be designed from two plasmons localized within the SRR gaps and interacting with each other through SRR arms. This interaction gives rise to a new asymmetric subradiant mode with the plasmons within two gaps being out of phase [see Fig.~\ref{fig:fig1}(b)]. Owing to their matching symmetries, this additional subradiant mode can efficiently couple to the asymmetric mode of SRRs when their frequencies are tuned [see Fig.~\ref{fig:fig1}(c)], resulting in a cascaded double Fano resonance in the far-field spectra~\cite{33,34,35}. Moreover, since the wavelength of graphene plasmons can be altered by changing the Fermi level, this enables dynamically controllable interaction between the two subradiant modes making one of them less pronounced. Direct numerical calculations of the far-field spectra shown in Fig.~\ref{fig:fig2}(a) confirm that the hybrid metasurface supports one superradiant (broad peak) and two subradiant resonances associated with the Fano resonances, one of them being highly tunable. Profiles of two subradiant asymmetric modes are shown in Fig.~\ref{fig:fig1}(b) for the case of their strong hybridization. Changing the frequency of the incident plane wave allows to see clearly the features of two resonances in near-field (see Supplementary Material~\cite{36}).


{\em Analytical approach.} The basic principle of the formation of the cascaded Fano resonances can be illustrated
by applying the coupled-mode theory~\cite{29}. In this approach, each mode of the metasurface is described by a complex amplitude ($S$, $A$, and $P$), corresponding to the bright symmetric and dark asymmetric modes of SRR, and a dark asymmetric graphene plasmonic mode, respectively. All modes are normalized in such a way that $|S|^2$ , $|A|^2$, $|P|^2$ are the energy densities. The amplitudes of the incident $E^+$ and reflected $E^-$ waves are also normalized such that $|E^{(+,-)} |^2$ represent the incoming and outgoing power fluxes. Then, the system dynamics is described by a set of coupled equations
\begin{equation}
\left \{
\begin{aligned} \label{eq:CMt}
\dot{S} &= (- i \omega_{S} - \gamma_S ) S + ig_1 A + i \varkappa_S E_{\text{in}}^{+}\:,\\
\dot{A} &= (- i \omega_{A} - \gamma_A) A + ig_1 S +  i g_2 P\:,\\
\dot{P} &= (- i \omega_{P} - \gamma_P) P + ig_2 A \:,
\end{aligned}
\right.
\end{equation}
where $\omega_m$ and $\gamma_m$ are bare frequencies and decay rates of the modes, $g_{1,2}$ characterize the interaction strengths, and $\varkappa_S$ determines coupling of the symmetric mode to an incident plane wave. In general, the total decay rate is composed of two parts $\gamma_{S,A,P} = \gamma^{\text{ohm}}_{S,A,P} +  \gamma^{\text{rad}}_{S,A,P}$, Ohmic (resistive) and radiative loss, respectively.
It is assumed that only the symmetric mode is driven by the time-harmonic incident field, 
while the asymmetric modes can be excited only indirectly through coupling. Additional constraints should be imposed on the coefficients to satisfy the energy conservation and time-reversal: $\gamma^{\text{rad}}_{A,P}=0$,
$|\varkappa_S| = \sqrt{\gamma^{\text{rad}}_{S}} $;
$ E_{\text{in}}^{+} = E_{1}^{+}$, where the numeric subscript (1 or 2) refers to the media surrounding the metasurface. For simplicity, the structure is assumed to be top-bottom symmetric, which neglects substrate effect justified by small optical contrast (the reflection coefficient without resonances $r= 0$) so
that $E_{1(2)}^{-} = E_{2(1)}^{+} + i  \varkappa_S S $. Complex reflection can then be found from expression $r= E_{1}^{-}/E_{1}^{+} = (i \varkappa_S S)/E_{1}^{+}$ and assumes the form of a "nested" Fano resonance:
\begin{equation*}
\label{eq:Ref}
r{=}{}\dfrac{i\varkappa_S^2}{\left[(\omega_{S}{-}i\gamma_S){-}\omega \right]{-}\dfrac{g_1^2}{\left[(\omega_{A}{-}i\gamma_A){-}\omega\right]{-}\dfrac{g_2^2}{\left[(\omega_{P}{-}i\gamma_P){-}\omega\right]}}},
\end{equation*}
which clearly reflects cascaded coupling of the plasmonic asymmetric mode to the radiative continuum via the asymmetric and symmetric modes of SRRs. The transmission coefficient can then be found from the equation $t = E_{2}^{-}/E_{1}^{+}= 1 +r$.

A change in the graphene chemical potential leads to variation of the spectral position of the plasmonic resonance $\omega_P$. Due to strong near-field coupling of two subradiant modes their interaction reveal anticrossing of associated Fano resonances, shown in Fig.~\ref{fig:fig2}(a). The analytical model nicely reproduces this behavior. In Fig.~\ref{fig:fig2}(a) transmission calculated with the use of COMSOL Multiphysics is superimposed with the dispersion curves found from the coupled-mode theory, with parameters recovered from the fitting. The spectral positions of the noninteracting modes are shown by dashed lines. Here the graphene plasmon branch is obtained from the standing-wave condition $2\pi/k_{\text{sp}}\approx g$, where $k_{\text{sp}}$ satisfies the dispersion of the p-polarized plasmons on the graphene layer on top of dielectric substrate $\varepsilon_2$: $\displaystyle{\frac{1}{\sqrt{k_{\text{sp}}^2-  k_0^2}}+\frac{ \varepsilon_2}{\sqrt{k_{\text{sp}}^2-\varepsilon_2 k_0^2}}= - i \frac{4\pi\sigma(\omega)}{\omega}}$. Interaction between the resonances leads to their hybridization with dispersion branches of the normal modes shown by dotted lines.

\begin{figure}[b]
\centering\includegraphics[width=1.0\linewidth]{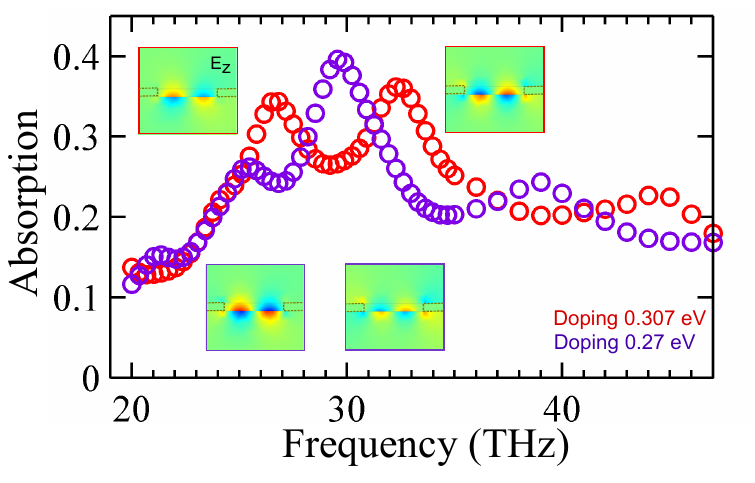}
\caption{(Color online) Numerically calculated absorption by the metasurface for graphene doping ${\mathcal E}_F = 0.27$ eV (purple circles) and ${\mathcal E}_F = 0.307$ eV (red circles). The insets show the field distributions in the SRR gap in the vicinity of two main absorption peaks associated with the excitation of Fano resonances.
}
\label{fig:fig3}
\end{figure}

Since the resonant frequency of graphene plasmons depends strongly on doping, the plasmonic mode intersects the subradiant SRR mode. Their interaction results in the characteristic avoided crossing behaviour with the frequency splitting between two modes. The results of the numerical and analytical calculations are plotted alongside in Fig.~\ref{fig:fig2}(a) and Fig.~\ref{fig:fig2}(b), respectively, where one can see that the tunable anti-crossing behavior described by these approaches agrees perfectly well.

It is interesting to mention that, in a sharp contrast to conventional tunable plasmonic resonances in graphene, the subradiant nature of the graphene resonance considered here allows not only tuning its spectral position, but also coupling  to a radiative continuum. Indeed, as demonstrated in Fig.~\ref{fig:fig2}, the intensity of the asymmetric plasmonic mode fades away as one moves away from the SRR asymmetric mode, evidencing the reduction of the radiative coupling caused by the suppression of the cascaded interaction between the modes. Therefore, the proposed metasurface offers an unprecedented control over its scattering characteristics, including spectral position, radiative coupling strength, and phase, unattainable in conventional designs of Fano-resonant metasurfaces. The ability to control radiative coupling can be of special interest for applications where the tunable near-field enhancement is a key requirement. To demonstrate the possibility of a control over strength of light-matter interaction, below we consider two examples: tunable absorption enhancement and nonlinear second-harmonic generation.

{\em Tunable absorption.} Frequency-dependent absorption is plotted in Fig.~\ref{fig:fig3} for two levels of graphene doping, with the insets showing the field distribution in the gap in the vicinity of two absorption peaks stemming from excitation of subradiant modes. The enhanced absorption due to excitation of hybrid graphene plasmon-quadrupole resonances clearly correlates with near-field enhancement. As expected, for the case of poorly interacting subradiant modes, corresponding to the case of lower Fermi energy ($\mathcal{E}_F=0.27$ eV, shown by purple circles in Fig.~\ref{fig:fig3}), the asymmetric plasmonic mode has smaller amplitude and the loss are not as prominent as for the asymmetric SRR resonance. For higher doping level ($\mathcal{E}_F=0.307$ eV, shown by red circles in Fig.~\ref{fig:fig3}) the two dark modes are strongly hybridized leading to a higher absorption rate by the tunable subradiant mode. This confirms that the system considered here can be promising for various applications where tunable strong field enhancement is in demand.

{\em Nonlinear effects.} Being a strongly nonlinear material, graphene is of a great interest for nanoscale nonlinear optics~\cite{39}, especially in combination with plasmonic effects~\cite{40,41,42}. Extreme surface confinement of electromagnetic fields due to excitation of plasmons is known to result in enhanced nonlinear response. Even stronger field confinement in hybrid metasurfaces proposed here can further boost the efficiency of various nonlinear optical effects~\cite{30,37,38}. In particular, the local field enhancement can be exploited for nonlinear frequency conversion. As the demonstration of enhanced nonlinear effects associated with the resonant excitation of subradiant plasmonic mode in graphene, here we study the second-harmonic generation (SHG) caused by a quadratic nonlocal nonlinearity of graphene.

The nonlinear surface current induced in graphene by the incident wave of the fundamental frequency (FF) $\omega$,
at the double frequency $2\omega$ has the form $ j_i(2\omega) = \sigma^{(2)}_{ijkl} (\omega) E^{\omega}_j \nabla_{k} E^{\omega}_l $, where $i,j,k,l$ are the Cartesian components. It depends on the gradient of FF electric field at the graphene surface, and it can be expressed through the second-order conductivity tensor~\cite{43},
\[
\sigma^{(2)}_{ijkl} (\omega) = \displaystyle{\frac{i e^3 V_F^2 } {8 \pi \hbar^2 \omega^3}} (5\delta_{ij}\delta_{kl} - 3\delta_{ik}\delta_{jl} +\delta_{il}\delta_{jk}),
\]
which is derived within the quasi-classical approach based on the Boltzmann equation~\cite{44,45,46,47}, applicable at low frequencies $\hbar \omega \leq \mathcal{E}_{F}$ corresponding to the plasmonic regime. In what follows, we neglect nonlinearity in metal.

The nonlinear response of the metasurface is simulated in COMSOL Multiphysics with the use of two coupled electromagnetic models assuming undepleted pump field. As the first step, we perform simulations at the fundamental (pump) wavelength 10.65 $\mu$m (matching one of the radiation lines of $\text{CO}_2$ laser) and stimulating excitation of graphene plasmon resonance in the SRR gap for the doping level $\mathcal{E}_F=0.31$ eV. Next, the nonlinear current induced at every point on the graphene surface for the second-harmonic (SH) frequency is calculated. This current is employed as a nonlinear source for the electromagnetic simulation and the SH fields generated by this current are obtained.

\begin{figure}[t]
\centering\includegraphics[width=1.0\linewidth]{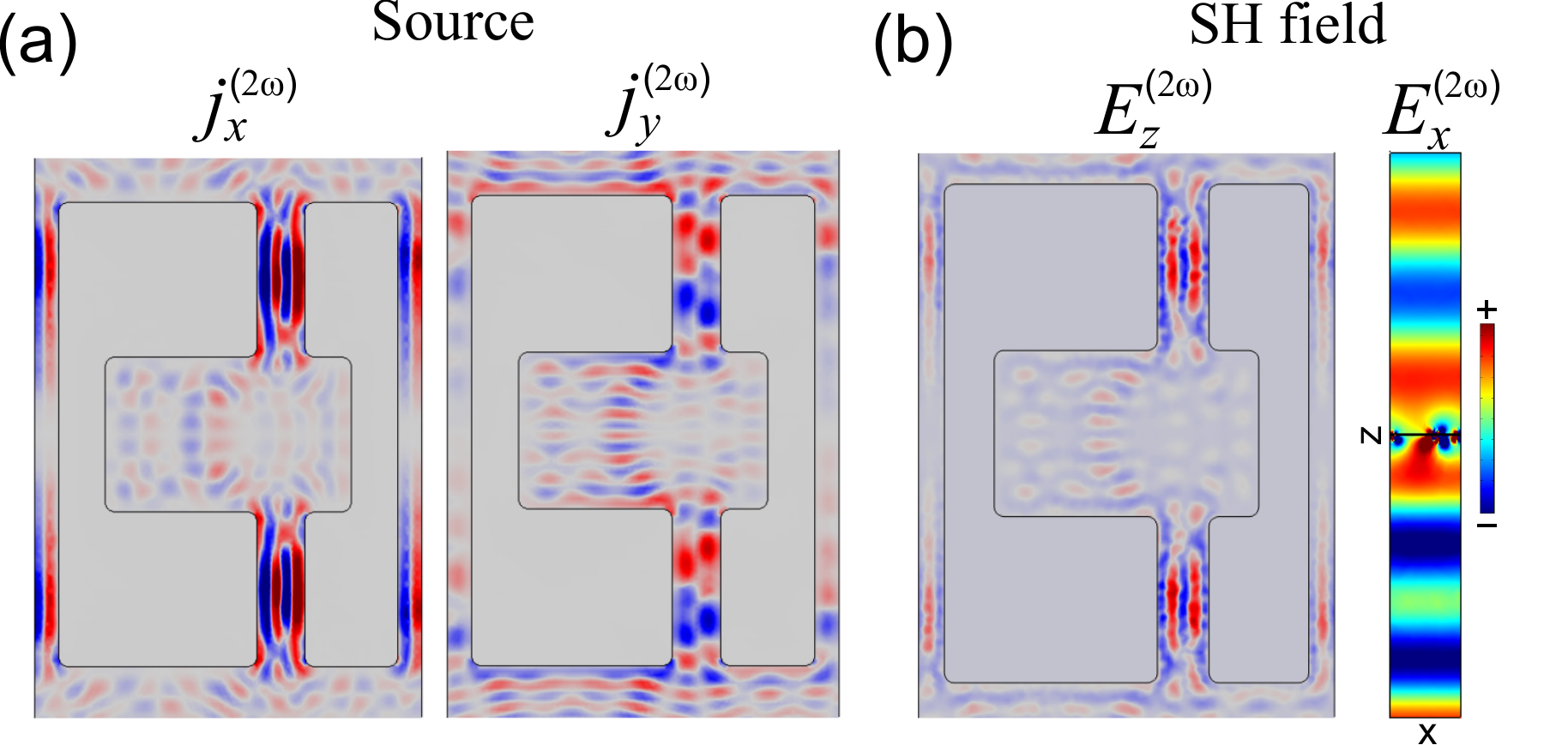}
\caption{(Color online)
(a) Nonlinear source and (b) SH fields generated by the hybrid metasurface; the $z$-component of the electric field is evaluated 40 nm above the graphene layer (left), and a side-view of the $x$-component evaluated in the middle of the SRR gap (right).
}
\label{fig:fig4}
\end{figure}

As expected, SHG predominantly arises from the gap, where the field of the asymmetric plasmonic mode is localized and strongly enhanced. The calculated nonlinear current generated by this subradiant mode is found to have the dominant $x$-component driving the entire metasurface at the SH frequency. As confirmed by numerical simulations shown in Fig.~\ref{fig:fig4}, this results in a non-vanishing nonlinear dipole moment aligned along $x$ axis within the SRR gaps, which leads to the generation of a $x$-polarized plane wave in the far-field region. It is interesting that the SH wave appears to be cross-polarized with respect to $y$-polarized incident plane wave at the fundamental frequency, indicating the nonlinear polarization conversion. We find that due to the strong field enhancement associated with two complimentary trapping mechanisms of electromagnetic field, namely the plasmonic confinement and a subradiant character of the plasmonic mode, the structure exhibits high efficiency of nonlinear conversion $\sim10^{-6}$ at the incident wave intensity of 1 MW/cm$^2$ even without matching to any resonances at the SH frequency. The animation illustrating time dependence of the SH field component $E_x$ radiated by the metasurface at the frequency of graphene plasmonic Fano resonance in the plane perpendicular to the SRR gap can be found in Supplementary Material~\cite{36}.

{\em Conclusions.}  We have proposed a novel type of graphene-based metasurfaces exhibiting cascaded Fano resonances originating from the coupling of standing graphene plasmons and the modes of structured metasurfaces. We have shown that such hybrid metasurfaces allow high tunability of the subradiant modes facilitated by controllable intermodal coupling. Such tunable interaction between subradiant modes, achieved though graphene doping, enables controlling the spectral position and lifetime of the Fano resonances. By utilizing two complimentary mechanisms of light confinement originating in the subwavelength localization and subradiant nature of the engineered graphene plasmonic modes, we have shown tunable and enhanced light-matter interaction. In particular, a significant enhancement of light absorption and nonlinear effects have been demonstrated. The proposed concept envisions tunable nonlinear response of either graphene or any nonlinear materials sputtered on top of the metasurfaces supporting cascaded Fano resonances. The demonstrated possibility to enhance and switch the second-harmonic generation by graphene gating can be used in various applications, e.g. for the efficient generation of short pulses of the second-harmonic field.

{\em Acknowledgements.} The authors acknowledge a support from the Australian Research Council. A part of research was carried out at the Center for Functional Nanomaterials of the Brookhaven National Laboratory supported by the U.S. Department of Energy, Office of Basic Energy Sciences under Contract No. DE-SC0012704


\end{document}